# Gate-Tunable Ambipolar Josephson Current in a Topological Insulator

Bomin Zhang[1,3], Xiaoda Liu[1,3], Junjie Qi[2,3], Ling-Jie Zhou[1], Deyi Zhuo[1], Han Tay[1], Hongtao Rong[1], Annie G. Wang[1], Zhiyuan Xi[1], Chao-Xing Liu[1], Chui-Zhen Chen[2], and Cui-Zu Chang[1]

[1]Department of Physics, The Pennsylvania State University, University Park, PA 16802, USA

[2]Institute for Advanced Study and School of Physical Science and Technology, Soochow University, Suzhou 215006, China

[3]These authors contributed equally: Bomin Zhang, Xiaoda Liu, and Junjie Qi

Corresponding authors: czchen@suda.edu.cn (C.-Z. Chen); cxc955@psu.edu (C.-Z. Chang).

**Abstract:** Dirac surface states in a topological insulator (TI) with proximity-induced superconductivity offer a promising platform for realizing topological superconductivity and Majorana physics. However, in TIs, the Josephson effect is usually observed in regimes where transport is dominated by either substantial bulk conduction channels or unipolar surface states. In this work, we demonstrate gate-tunable ambipolar Josephson current in lateral Josephson junction (JJ) devices based on bulk-insulating $(Bi,Sb)_2Te_3$ thin films grown by molecular beam epitaxy (MBE). For thinner films, the supercurrent exhibits pronounced gate-tunable ambipolar behavior and is significantly suppressed as the chemical potential approaches the Dirac point, yet persists across it. In contrast, thicker films exhibit a much weaker ambipolar response. Moreover, we find that the supercurrent becomes significantly less resilient to external magnetic fields when the chemical potential is tuned near the Dirac point in both thickness regimes. Our numerical simulations demonstrate the ambipolar behavior of these TI JJ devices and attribute the asymmetric supercurrent observed in thicker TI films to the coexistence of Dirac surface states and bulk conduction channels. The **demonstration of gate-tunable ambipolar Josephson transport in MBE-grown TI films paves the way for realizing Dirac-surface-state-mediated topological superconductivity and establishes a foundation for future exploration of electrically tunable Majorana modes.**

**Main text:** Over the past two decades, the search for the topological superconducting (TSC) phase has attracted intense attention because of its deep connection to fundamental physics and its potential application in topological quantum computations [1,2]. The TSC phase has been proposed to emerge in hybrid structures where one-dimensional (1D) or two-dimensional (2D) electronic systems with spin-split bands, caused by strong spin-orbit coupling (SOC), are proximally coupled to a conventional *s*-wave superconductor [3-5]. TSC possesses a full superconducting gap in the bulk but shows topologically protected zero-energy states on its boundaries or surfaces, or bound to the vortex cores imposed by an external magnetic field. These zero-energy states, consisting of an equal mixture of electrons and holes, are analogous to the self-conjugate property of Majorana fermions in particle physics (i.e., a particle that is its own antiparticle) [6], and are thus referred to as Majorana zero modes (MZMs).

Topological insulators (TIs) possess inherent linearly-dispersed Dirac surface states, which are induced by strong SOC (Refs. [7,8]), making them an ideal platform for pursuing MZMs (Refs. [9-11]). Moreover, the large bulk band gap of a few hundred meV in TIs isolates the Dirac surface states from bulk excitations [12], providing more flexibility to realize MZMs. To probe MZMs in TI/superconductor hybrid structures and pave the way for future device applications, Josephson junction (JJ) devices and superconducting quantum interference devices (SQUIDs) are essential [13-16]. These devices are "gate-friendly", i.e., the chemical potential of the TI film can be readily tuned by a gate to access its Dirac surface states. The JJ device is formed by connecting two conventional *s*-wave superconductors through a weak link of TI films, while the SQUID device is formed by a superconducting loop containing JJ devices. In both JJ and SQUID devices, the current-phase relation (CPR) can be used to investigate mesoscopic proximity-induced superconductivity within the material of the junction region [17,18]. TI-based JJ devices have revealed signatures of 4π-periodicity [19-21], phase-coherent transport through edge states [21-23], highly skewed CPR (Ref. [24]), SOC-induced $\varphi_0$ junction [25], and nano-SQUID driven by an in-plane magnetic field [26]. To date, several studies have claimed that JJ devices are realized based on bulk-insulating TIs (Ref. [21,22,24,27-33]), but none have demonstrated ambipolar Josephson current, a supercurrent in a single JJ device that can be carried by either electrons or holes, as observed in graphene [34]. The key impediment is the inability to cleanly separate the proximity-induced superconductivity in the Dirac surface states from bulk conduction channels, which compromises TI-based Majorana qubit schemes. Demonstrating a gate-tunable ambipolar Josephson current in a single device would overcome this bottleneck and yield an unambiguous signature of Dirac-surface-state-dominated TSC.

In this work, we employ molecular beam epitaxy (MBE) to grow $d$ quintuple layer (QL) $(Bi,Sb)_2Te_3$ films on heat-treated insulating $SrTiO_3$(111) substrates. We fabricate lateral JJ and SQUID devices based on these TI films. For the $d = 5$ JJ device (denoted as JJ-1), we observe gate-tunable ambipolar Josephson current. In the crossover regime between the conduction and valence bands, the normal-state resistance $R_n$ exhibits a sharp peak, where both the critical current $I_c$ and the $I_cR_n$ product are suppressed. For the $d = 15$ JJ device (JJ-2), the ambipolar Josephson transport behavior is significantly reduced, while an asymmetry between $p$- and $n$-type carrier regimes is observed. Specifically, the $p$-type side exhibits a substantially larger $I_c$ compared to the $n$-type side. For both samples, supercurrent exhibits reduced magnetic resilience in the bulk-insulating regime relative to the metallic regime, but persists under finite magnetic fields. Our numerical calculations reveal that the asymmetric supercurrent observed in the JJ-2 device is a precursor to ambipolar transport, the onset of which is suppressed by the coexistence of gate-insensitive bulk conduction channels and top-layer Dirac surface states originating from structural inversion asymmetry (SIA) between the top and bottom surfaces.

The chemical potential of the $(Bi,Sb)_2Te_3$ films is roughly tuned by optimizing the Bi/Sb ratio during the MBE growth and further finely adjusted using a SrTiO$_3$-assisted bottom gate during electrical transport measurements. The superconducting Nb patterns that define the junction area are deposited in a DC sputtering chamber. The $(Bi,Sb)_2Te_3$ films, except in the junction and Nb electrode areas, are precisely etched away by argon plasma milling to enhance gating efficiency. Electrical transport measurements are conducted using a dilution refrigerator (Bluefors LD250, 10mK, 9-1-1 T). More details about the MBE growth, device fabrication, and electrical transport measurements can be found in Methods.

Figure 1a shows a schematic of our JJ devices, which include an MBE-grown $(Bi,Sb)_2Te_3$ film mesa on an insulating SrTiO$_3$ substrate, two isolated Nb electrode patterns deposited on the two ends of the $(Bi,Sb)_2Te_3$ film, and a global SrTiO$_3$-assisted back gate. The junction area is $L \times W$ = 20 nm × 2 μm. A scanning electron microscope (SEM) image of our JJ device is shown in Fig. 1b. We first conduct electrical transport measurements on the JJ-1 device with $d$ = 5 to determine the position of the Dirac point in the TI layer. The normal state resistance $R_n$ exhibits a sharp peak at the charge neutral point, i.e., $V_g = V_g^0$ (Fig. 1e), corresponding to the Dirac point of the 5 QL $(Bi,Sb)_2Te_3$ film. Figure 1c shows three selected voltage-current $V$-$I$ curves at different gate biases ($V_g$ - $V_g^0$). The current source measurements are performed by sweeping $I$ from zero to a finite bias. By combining zero-up and zero-down sweeps, we can avoid hysteresis from heating effects and underdamped Josephson current [35,36]. For ($V_g$ - $V_g^0$) = 0 V, i.e., near the Dirac point, the Josephson effect is observed with a critical current $I_c \sim 40$ nA, identified by two sharp voltage-switching transitions. For $(V_g - V_g^0) = -40$ V, i.e., the *p*-type regime, the $I_c$ value increases to ~200 nA. For $(V_g - V_g^0) = +40$ V, i.e., the *n*-type regime, the $I_c$ value is ~210 nA.

Figure 1d shows a color plot of the differential resistance d$V$/d$I$ as a function of $I$ and $(V_g - V_g^0)$ at an external magnetic field $B_z = 0$ T. The red lines indicate the current biases at which the junction transitions from the superconducting to normal states, corresponding to the values of $I_c$ at different $V_g$. For $-40$ V $\leq (V_g - V_g^0) \leq -10$ V, $I_c$ slightly decreases as $(V_g - V_g^0)$ increases, with $I_c \sim 180$ nA at $(V_g - V_g^0) = -10$ V. For $+7$ V $\leq (V_g - V_g^0) \leq +40$ V, $I_c$ remains nearly constant. However, for $-10$ V $< (V_g - V_g^0) < +7$ V, where the chemical potential shifts from the bulk conduction bands (*n*-type), through the Dirac surface states, and into the bulk valence bands (*p*-type), $I_c$ first decreases and then increases. We note that $I_c$ reaches its minimum value of ~40 nA at $(V_g - V_g^0) = 0$ V.

To estimate the strength of the Josephson effect in the JJ-1 device, we plot the $(V_g - V_g^0)$ dependence of the product of the critical current and the normal state resistance $I_cR_n$ (Fig. 1e). We find that the value of $I_cR_n$ remains nearly constant for $-40$ V $\leq (V_g - V_g^0) \leq -10$ V and $+8$ V $\leq (V_g - V_g^0) \leq +40$ V and reaches a minimum at $(V_g - V_g^0) = 0$ V, indicating the Josephson effect weakens as the chemical potential approaches the Dirac point. In the JJ-1 device, $I_cR_n$ ranges from 12 µV to 35 µV, about an order larger than the *ex-situ* fabricated $(Bi,Sb)_2Te_3$-based JJ devices (2~4 µV) in prior studies [29], though still slightly lower than the $(Bi,Sb)_2Te_3$-based JJ devices achieved with *in-situ* formation (~53 µV) (Ref.[37]) and self-epitaxy method (~60 µV) (Refs.[38,39]). In the JJ-5 device with $d = 15$, we also achieve $I_cR_n \sim 50$µV (Supplementary Fig. 5b). Next, we assess the superconducting proximity effect between the Nb electrodes and the 5 QL $(Bi,Sb)_2Te_3$ film by extracting the induced superconducting gap $\Delta$ from the voltage bias $V$-dependent differential conductance d$I$/d$V$ (Supplementary Fig. 3). The extracted $\Delta$ values are found to be ~20 µeV at ($V_g$

- $V_g^0$) = -30 V (i.e., the *p*-type regime) and ~12.5 μeV at ($V_g$ - $V_g^0$) = 0 V (i.e., near the Dirac point). These results indicate that the proximity-induced superconductivity in the 5 QL (Bi,Sb)$_2$Te$_3$ film is suppressed when the chemical potential approaches the Dirac point, consistent with prior studies on JJ devices based on Dirac materials [21-24,26-29,32-34].

Next, we further evaluate the JJ-1 device by quantitatively calculating a few characteristic parameters. The $eI_cR_n/\Delta$ ratio is ~1 across the entire ($V_g$ - $V_g^0$) range, where *e* is the electron charge. This ratio, being on the same order as the theoretical limit of $\frac{\pi}{2}$ (Ref. [40]), suggests that the JJ-1 device has few defects across the Nb/(Bi,Sb)$_2$Te$_3$ interface. Based on a Hall bar fabricated on the same substrate, we extract a superconducting coherence length $\xi_N$ ~ 452 nm and a mean free path $l_{mfp}$ ~ 24.8 nm in the 5 QL (Bi,Sb)$_2$Te$_3$ film (Supplementary Fig. 2). There, if we assume supercurrent transport occurs entirely through the Dirac surface states, the JJ-1 device satisfies $l_{mfp}$ ~ $L \ll \xi_N$, indicating a short junction in the intermediate region between ballistic and diffusive transport. More details of these parameter estimations can be found in Supplementary Information. Two more JJ devices with *d* = 5 (JJ-3 and JJ-4), fabricated using the same recipe, exhibit similar gate-tunable ambipolar Josephson current behavior (Supplementary Fig. 5).

To study the dependence of the Josephson effect on *d*, we fabricate a JJ device with *d* = 15 (JJ-2) using the same recipe and perform transport measurements on this device (Fig. 2). For -80 V ≤ ($V_g$ - $V_g^0$) ≤ 0 V, i.e., the *p*-type regime, the value $R_n$ increases monotonically as ($V_g$ - $V_g^0$) increases. $I_c$ reaches a maximum of ~400 nA at ($V_g$ - $V_g^0$) = -10 V, where the $I_cR_n$ product also attains its highest value of ~35 μV. For 0 V ≤ ($V_g$ - $V_g^0$) ≤ 40 V, i.e., the *n*-type regime, $R_n$ remains nearly constant, while both $I_c$ and $I_cR_n$ show a slight increase as ($V_g$ - $V_g^0$) increases. At ($V_g$ - $V_g^0$) = 0 V, i.e., near the Dirac point, $I_c$ ~40 nA and $I_cR_n$ ~7 μV, both are the minimum values across the entire ($V_g$ - $V_g^0$) regime (Figs. 2a and 2b). Both $I_c$ and $I_cR_n$ values are comparable for JJ-1 and JJ-2 near the Dirac point. We attribute this observation to the high material quality of our MBE-grown (Bi,Sb)$_2$Te$_3$ films. Unlike the gate-tunable ambipolar Josephson current observed in the JJ-1 device, the JJ-2 device exhibits a much weaker ambipolar Josephson current with varying $V_g$, i.e., $I_c$ changes significantly in the *p*-type regime but remains almost constant in the *n*-type regime. Similar behaviors are observed in two more JJ devices with *d* =15 (JJ-5 and JJ-6) (Supplementary Fig. 6).

In the JJ-2 device, multiple resonant resistance peaks are observed in the color plot of d$V$/d$I$ for $(V_g - V_g^0) \leq -15$ V (Fig. 2a), which are characteristic of mesoscopic features in JJ devices and are typically associated with multiple Andreev reflections (MAR) and Andreev-bound states [41]. Figures 2c to 2e show *V*-dependent differential conductance d$I$/d$V$ measured at different $(V_g - V_g^0)$. At $(V_g - V_g^0) = -22$ V, i.e., the *p*-type regime, several d$I$/d$V$ peaks are observed (Fig. 2c), originating from the MAR process. Through a linear fit of these d$I$/d$V$ peaks positions (Fig. 2c inset), the induced superconducting gap $\Delta$ is estimated to be ~30 μeV, giving $eI_cR_n/\Delta$ ~ 0.6. At $(V_g - V_g^0) =$ +10 V, i.e., near the Dirac point, only two pairs of d$I$/d$V$ peaks corresponding to $\Delta$ and $2\Delta$ are observed, with $\Delta$ ~10 μeV (Fig. 2e). This indicates a suppressed proximity effect at low carrier density, consistent with the phenomenon observed in the JJ-1 device. The $eI_cR_n/\Delta$ ratio is ~ 0.8 at $(V_g - V_g^0) =$ +10 V, slightly larger than that at $(V_g - V_g^0) = -22$ V.

Remarkably, at $(V_g - V_g^0) = -22$ V, the peaks observed at the lowest *V* correspond to the index $n = 6$ of MAR. This demonstrates 6 coherent Andreev reflection processes in the junction area, suggesting a ballistic transport in the TI layer of the JJ-2 device, which also agrees well with the $eI_cR_n/\Delta$ ratio of ~1 in this JJ device. At $(V_g - V_g^0) = -10$ V, these MAR peaks disappear (Fig. 2d), presumably due to the DC voltage bias measurements setup: the filter resistance used in our measurements is ~3.5 kΩ, while the normal state resistance $R_n$ of the JJ-2 device is a few hundred

Ω (Fig. 2b). Due to their series connection, electrical transport measurements on the JJ-2 device favor the current source configuration, making it difficult to resolve MAR peaks [42,43]. More discussion can be found in Supplementary Information.

To further examine the proximity-induced superconductivity in our JJ devices, we measure the dependence of $I_c$ on an out-of-plane magnetic field $B_z$ in the JJ-1 and JJ-2 devices. Figures 3a to 3c show the Fraunhofer patterns of the JJ-1 device measured at $(V_g - V_g^0)$ = -40 V (i.e., the *p*-type regime), 0 V (i.e., near the Dirac point), and +40 V (i.e., the *n*-type regime). The critical supercurrent $I_c$ of a JJ device with a homogeneous current distribution can be calculated using the equation:

$$I_c(B_z) = I_c(B_z = 0)|\frac{\sin(\pi\Phi/\Phi_0)}{\pi\Phi/\Phi_0}| \qquad (1)$$

where $\Phi_0 = h/(2e)$ is the magnetic flux quantum and $\Phi$ is the net flux passing through the junction area. An average modulation period is estimated to be $B_0 \sim 0.8$ mT from the side lobes. Given the width of the junction area $W \sim 2$ μm, the effective length of the junction area is calculated as $L_{eff} = \Phi_0/WB_0 \sim 1$ μm. This effective length is significantly larger than the length of the fabricated JJ device length $L$. Similar behavior has been observed in planar JJ devices based on 2D electron gases [44-46] and topological materials [28,39,47], and is attributed to the London penetration of the Nb electrodes and flux focusing induced by the Meissner effect [48]. The amplitude of the side lobes in the Fraunhofer patterns does not decay monotonically as $B_z$ increases, and the lobe periods are nonuniform (Figs. 3a to 3c). This behavior, known as anomalous Fraunhofer patterns, is usually attributed to the nonuniform current distribution or variations in the effective channel length within the junction area [28,49]. The SEM image of the JJ device on a TI film reveals that the Nb electrodes have rough edges (Fig. 1b). Such morphology, together with flux focusing effect, can cause the effective length $L_{eff}$ to vary with position $x$ across the junction area [28], where $-W/2 \leq x \leq +W/2$.

Next, we analyze the anomalous Fraunhofer patterns of the JJ-1 device at different ($V_g - V_g^0$). For ($V_g - V_g^0$) = -40 V (i.e., the *p*-type regime) and ($V_g - V_g^0$) = +40 V (i.e., the *n*-type regime), the d$V$/d$I$ peaks associated with superconducting-to-normal state transitions remain sharp across the first four side lobes (Figs. 3a and 3c). However, for ($V_g - V_g^0$) = 0 V, i.e., near the Dirac point, the d$V$/d$I$ switching peaks are broadened in all side lobes, with the second and third side lobes being unresolvable compared to heavily-doped regime (Fig. 3b). This observation suggests weaker magnetic field resilience at the Dirac point, where the Josephson effect is suppressed due to a reduced proximity-induced superconducting gap Δ in the TI layer (Supplementary Fig. 3). Nevertheless, the appearance of the periodic Fraunhofer patterns at finite $B_z$ confirms that the phase-coherent transport persists in the JJ-1 device, even at the Dirac point. We note that proximity-induced superconductivity in the TI layer, when its chemical potential crosses the Dirac point, opens opportunities to explore TSC and Majorana physics in bulk-insulating TI/superconductor [9-11] and quantum anomalous Hall insulator/superconductor hybrid structures [50-52]. Figures 3d to 3f show the Fraunhofer patterns of the JJ-2 device at different ($V_g - V_g^0$), where anomalous Fraunhofer patterns are also observed. At ($V_g - V_g^0$) = 0 V, i.e., near the Dirac point, the supercurrent $I$ similarly exhibits reduced resilience to $B_z$ (Fig. 3f).

To support our experimental results, we perform numerical simulations of the critical Josephson current in a TI-based JJ using the recursive Green's function method [42-44]. Our aim here is to understand the ambipolar behaviors of Josephson currents in TI films, and particularly the strong asymmetry between *n*- and *p*-type regimes observed in the JJ-2 device with $d$ = 15, as compared to the JJ-1 device with $d$ = 5. We first consider a 3D TI model that incorporates a SIA between the top and bottom surfaces, with the top layer coupled with a conventional *s*-wave superconductor (Methods). The SIA originates from the work function difference between the TI film and the SrTiO$_3$(111) substrate, which results in a gradual change in the chemical potential of the TI film along its normal direction [12]. Next, we calculate the critical current, $I_c$, as a function of the carrier density, $(n - n_0)$, using the same geometric configuration employed in our experiments (Methods). Here $n_0$ denotes the carrier density at the Dirac point of the top surface, which corresponds to $V_g^0$ in the experiment. Figure 4a shows ambipolar Josephson current in a TI-based JJ device with $d = 5$, closely matching the experimental results of the JJ-1 device (Fig. 1d). Specifically, the Josephson critical current $I_c$ increases as the carrier density $n_0$ increases in both $p$-type (i.e., $-6 \times 10^{12}$ cm$^{-2} \leq n - n_0 \leq 0$) and $n$-type (i.e., $0 \leq n - n_0 \leq 5 \times 10^{12}$ cm$^{-2}$) regimes. It reaches a maximum value of $I_c \sim 60e\Delta/\hbar$, with a pronounced minimum $I_c \sim 20e\Delta/\hbar$ at the Dirac point (i.e., $n = n_0$). As noted above, for the JJ-1 device, the extracted induced superconducting gap $\Delta$ ranges from ~12.5 μeV at $(V_g - V_g^0) = 0$ V to ~20 μeV at $(V_g - V_g^0) = -30$ V. By assuming $\Delta = 12.5$ μeV, the simulated theoretical value $I_c \sim 60$ nA at the Dirac point is of the same order as the measured $I_c$ at the Dirac point of the JJ-1 device (Fig. 1d), indicating that transport in the device is in the ballistic regime.

To further investigate the dependence of the Josephson effect on $d$, we simulate the critical Josephson current for a TI-based JJ device with $d = 15$, corresponding to the JJ-2 device in our experiments (Fig. 2). Unlike the gate tunable ambipolar Josephson current observed in the $d = 5$ device, the $d = 15$ device exhibits asymmetric behavior of $I_c$ compared to the Dirac point (Fig. 4b). Specifically, $I_c$ varies significantly in the $p$-type regime but exhibits a slight increase as $(n - n_0)$ increases in the $n$-type regime, closely matching the experimental results of the JJ-2 device (Fig. 2a). The strongly suppressed ambipolar Josephson current in the $d = 15$ device can be attributed to the coexistence of Dirac surface states and bulk conduction channels. For the $d = 15$ device, the top-layer Dirac surface states (Fig. 4d) overlap with the bulk conduction bands due to a pronounced band shift induced by the large chemical potential difference between the top and bottom surfaces. Consequently, when the gate drives the system into the $n$-type, these bulk conduction channels are only weakly tunable by $V_g$, leading to a slow increase in the critical supercurrent $I_c$. In contrast, for the $d = 5$ device, the SIA is weak [12]. The Dirac surface states are slightly shifted (Fig. 4c). Therefore, our numerical simulation provides a scenario to understand the asymmetric behavior of Josephson currents observed in the JJ-2 device, which can be interpreted as a precursor to ambipolar transport, hindered by the presence of finite bulk conduction channels.

We further extend the TI-based JJ fabrication technique to realize symmetric SQUID devices (SQUID-1 and SQUID-2 in Supplementary Figs. 7 and 8). The SQUID-1 device is fabricated on a 10 QL $(Bi,Sb)_2Te_3$ film, while the SQUID-2 device uses a 15 QL $(Bi,Sb)_2Te_3$ film. Each device consists of two identical JJs connected in parallel, with the chemical potential of the TI film tuned by a $SrTiO_3$-assisted global $V_g$. The CPR in the junctions is precisely driven by the SQUID loop area, effectively minimizing the effect of field penetration into the superconductor electrodes and anomalous Fraunhofer patterns. Prior studies on TI-based SQUID devices have focused on the CPR in the junction, including the emergence of higher-order harmonics [24], possible $4\pi$-periodicity [53,54], and phase control for the detection of zero-bias conductance peaks [55]. The gate-tunable chemical potential offers an additional degree of control, enabling switching between topological trivial and nontrivial regimes and thus facilitating exploration of the TSC.

In our experiments, both SQUID-1 and SQUID-2 exhibit $V_g$-dependent $I_c$. SQUID-1 demonstrates ambipolar supercurrent behavior, whereas SQUID-2 exhibits weaker gate tunability and remains in the $p$-type regime even at the highest applied $V_g$. In both devices, the observed modulation periods correspond to a $2\pi$–periodic CPR (Supplementary Information). Moreover, magnetic flux-driven oscillations in both devices exhibit a sinusoidal shape that remains unchanged across the entire $V_g$ range applied (Supplementary Figs. 7d to 7f and 8b to 8e). The skewed CPR is absent in our device, which is reported as the signature of ballistic transport [56] and predicted to arise from surface-state-mediated transport [24,53,54]. Given that numerical results suggest the presence of mixing between bulk and surface states near the Dirac point of the TI film with $d$ =15, our observation indicates that supercurrent transport on pure surface states has not yet been achieved in these SQUID devices.

To summarize, we observe gate-tunable ambipolar Josephson current in JJ devices based on MBE-grown $(Bi,Sb)_2Te_3$ TI thin films. Although proximity-induced superconductivity weakens as the chemical potential is tuned into the bulk gap, a finite Josephson current persists at the Dirac point. The supercurrent carried by the TI Dirac surface states exhibits reduced resilience to external magnetic fields compared to the bulk conduction channels, reflecting the smaller induced gap at low carrier densities. We observe that, as the TI film thickness increases, the ambipolar Josephson current becomes asymmetric between the *n*- and *p*-type regimes. Tight-binding model calculations provide a theoretical scenario to understand this asymmetry and suggest that bulk conduction bands originating from band bending impede ambipolar Josephson current due to the presence of finite bulk conduction states. Our results demonstrate that JJ devices exhibiting ambipolar Josephson current provide a promising platform for investigating proximity-induced superconductivity in Dirac surface states. This work paves the way for exploring TSC and Majorana physics, as well as advancing scalable topological quantum computations based on MBE-grown TI films.

**Methods**

**MBE growth**

The TI films [i.e., $(Bi_{1-x}Sb_x)_2Te_3$ with $x$=0.25] with different thicknesses used in this work are grown on heat-treated 0.5 mm thick SrTiO$_3$(111) substrates in a commercial MBE chamber (Omicron Lab10) with a vacuum better than ~2 × 10$^{-10}$ mbar. Before the growth of the TI films, the 3 mm × 10 mm SrTiO$_3$(111) substrates are first outgassed at ~600 ℃ for 1 hour. Next, high-purity Bi (99.9999%), Sb (99.9999%), and Te (99.9999%) are evaporated from Knudsen effusion cells. During the MBE growth, the SrTiO$_3$(111) substrate is maintained at ~230 ℃. The Te/(Bi + Sb) flux ratio is greater than ~10 to prevent Te deficiency in the TI films. The Bi/Sb ratio is optimized to tune the chemical potential of the TI layer near the charge neutral point. The growth rate of the TI layer is ~0.2 QL per minute. No capping layer is used before the TI films are taken out of the MBE chamber.

**Nanofabrication of the JJ and SQUID devices**

The JJ and SQUID devices are fabricated on our MBE-grown TI films using electron-beam lithography. The electron-beam energy is ~100 keV. To prevent contamination and degradation in ambient conditions, the TI films are coated with PMMA 950 A3 resist as soon as they are removed from the MBE chamber. The TI film with the resist is first baked at ~180 °C for 15 minutes. Next, we employ electron-beam lithography (EBL) to define the Nb electrode patterns. After that, we clean the resist residue in an oxygen plasma asher. Finally, we use sputtering to deposit 2 nm Ti /40nm Nb as the electrodes. The Nb strips have a width $W$ of ~2 μm and a gap length $L$ of 20 ~ 40 nm. Following the liftoff process, the device is re-coated with PMMA 950 A3 resist. The resist is then left to stabilize at room temperature in a vacuum metal desiccator for 24 hours. Finally, except in the junction and electrode areas, the TI films are precisely etched away using EBL patterning followed by argon plasma milling.

**Electrical transport measurements**

Electrical transport measurements are conducted in a dilution refrigerator (Bluefors LD250, 10 mK, 9-1-1 T). The contacts for electrical transport measurements are made by pressing indium spheres on Nb electrodes. The bottom gate is prepared by flattening the indium dots on the back side of the SrTiO$_3$(111) substrates. The bottom gate voltage $V_g$ is applied using a Keithley 2450 source meter. The electrical current is applied using a Yokogawa GS200 DC source meter. The DC voltage signal across the JJ or SQUID devices is amplified using a LI-75A NF Corporation preamplifier and then measured with either a Keithley 2182A nanovolt meter or a Keysight 34461A digit multimeter. For two- or three-terminal measurements, the differential resistance d$V$/d$I$ is corrected by subtracting the series resistance $R_{in}$ from the RC filter in our dilution refrigerator.

**Theoretical calculations**

We start from a 3D TI model that incorporates a structure inversion asymmetry between the top and bottom surfaces,

$$H_T = H_{TI}(\boldsymbol{k}) + V(z) \tag{2}$$

where the low-energy effective TI Hamiltonian on the cubic lattice with lattice constant $a = 1$ nm is given by $H_{TI}(\boldsymbol{k}) = \sum_{i=0}^{3} d_i(\boldsymbol{k})\Gamma_i$ (Ref. [57]), with $d_i(\boldsymbol{k}) = A_i \sin(k_i a)/a$ and $d_0(\boldsymbol{k}) = M_0 - \sum_{i=0}^{3} 2B_i[1 - \cos(k_i a)]/a^2$. $V(z)$ is the potential distribution along the $z$-direction arising from the structural inversion asymmetry inherent to the film's growth on a substrate. The Dirac matrices $\Gamma$ are defined as $\Gamma_{(0,1,2,3)} = (\sigma_0 \otimes \tau_3, \sigma_1 \otimes \tau_1, \sigma_2 \otimes \tau_1, \sigma_3 \otimes \tau_1)$, with $\sigma_i$ and $\tau_i$ representing spin and orbital Pauli matrices, respectively. Here, the TI is topologically nontrivial when $M_0 B_i > 0$, and the $H_{TI}(\boldsymbol{k})$ describes a TI. We set parameters as $M_0 = 0.05$ eV, $A_{1,2,3} = 0.1$ eV·nm, $B_{1,2} = 0.025$ eV·nm$^2$, and $B_3 = 0.05$ eV·nm$^2$. We assume a linear potential profile $V(z) = V_0 \frac{z-1}{d-1}$ with $V_0 = 0.04$ eV for $d = 5$ and $V_0 = 0.06$ eV for $d = 15$ sample.

To incorporate superconducting proximity effects, we model the left and right superconducting electrodes as conventional s-wave superconductors. The Bogoliubov-de Gennes (BdG) Hamiltonian in Nambu space is expressed as

$$H_{BdG}(\mathbf{k}) = \begin{pmatrix} H_T(\mathbf{k}) - \mu_S & \Delta_{L(R)} \\ \Delta_{L(R)}^\dagger & -H_T^*(-\mathbf{k}) + \mu_S \end{pmatrix}, \quad (3)$$

where $H_T(\mathbf{k})$ is the Hamiltonian in Eq. (2) and $\mu_S$ is the chemical potential in the superconducting leads. The induced pair potential is given by $\Delta_{L(R)} = i\Delta e^{i\varphi_{L(R)}}\sigma_2 \otimes \tau_0$, with $\varphi_{L(R)}$ denoting the superconducting phases of the left (right) electrode. The pairing term is introduced exclusively on the top surface of 3D TIs ($z = 1$). The chemical potentials of the superconducting region are fixed at $\mu_S = 0.03$ eV for $d = 5$ and $-0.03$ eV for $d = 15$.

In our numerical simulations, we discretize the model Hamiltonian $H_{BdG}$ onto a cubic lattice with lattice constant $a = 1$ nm to construct a real-space tight-binding model. For computational efficiency, we adopt periodic boundary conditions along the $y$-axis, which makes $k_y$ a good quantum number with discretized values $k_y^m = 2\pi m/W$ in the first Brillouin zone. Here $m = 0, \cdots, (N_y - 1)$ and $W = N_y a$ is the sample width. The Josephson current $I_s$ flowing along the $x$-axis can be calculated using the recursive Green's function method [58-60],

$$I_s = -\frac{iek_B T}{\hbar} \sum_n \sum_{m=0}^{N_y-1} \text{Tr}\left[T_x G_{i+\delta_x, i}(i\omega_n, k_y^m) - T_x^\dagger G_{i, i+\delta_x}(i\omega_n, k_y^m)\right] \quad (4)$$

where $e$ is the elementary charge, $k_B$ is the Boltzmann constant, $\hbar$ is the reduced Planck constant, $\delta_x$ is the unit vector along the $x$-axis, and $T$ is the temperature, chosen as $T = \Delta/20$. The hopping term $T_x$ is given by

$$T_x = \begin{pmatrix} t_x & 0 \\ 0 & t_x^* \end{pmatrix} \tag{5}$$

with $t_x = -iA_1/2\sigma_1 \otimes \tau_1 + B_1 \sigma_0 \otimes \tau_3$ (6)

where $A_1, B_1$ are model parameters of $H_{BdG}$, and $\sigma_i$ and $\tau_i$ are spin and orbital Pauli matrices, respectively. The Matsubara's Green function is defined as

$$G(i\omega_n, k_y^m) = (i\omega_n - H_{BdG} - \Sigma_L - \Sigma_R)^{-1} \tag{7}$$

where $\omega_n = (2n+1)\pi k_B T$ is the Matsubara frequency, and the first summation is performed over all integers $n \in \mathbb{Z}$. Here, $\Sigma_{L(R)}$ represents the self-energy due to coupling between the central region and the left (right) superconducting Nb electrode, which can be calculated numerically. The critical Josephson current is defined as

$$I_c = \max_{0 \leq \varphi < 2\pi} I_s(\varphi) \tag{8}$$

where $\varphi = \varphi_R - \varphi_L$ is the superconducting phase difference. In our numerical simulations, we consider a JJ device with dimensions $L \times W$ = 20 nm × 2 μm, consistent with the geometry in our experiments.

**Acknowledgments:** This work is primarily supported by the NSF grant (DMR-2241327), including MBE growth and dilution transport measurements. The device fabrication and the sample characterization are supported by the DOE grant (DE-SC0023113). The PPMS measurements are supported by the NSF-CAREER award (DMR-1847811). C. Z. Chang acknowledges the support from the Gordon and Betty Moore Foundation's EPiQS Initiative (GBMF9063 to C. -Z. Chang). C.-Z. Chen acknowledges the support from the National Key R&D Program of China Grant (2022YFA1403700), the Natural Science Foundation of Jiangsu Province Grant (BK20230066), and the Jiangsu Shuang Chuang Project (JSSCTD202209). J. Q.

acknowledges the support from the National Natural Science Foundation of China (Grant No.12204053).

**Author Contributions:** C. -Z. Chang conceived and supervised the experiment. B.Z., X.L., L.-J.Z., A.G.W., and Z.X. fabricated all JJ and SQUID devices and performed electrical transport measurements. D.Z., H.T., and H.R. performed the MBE growth. J.Q., C.-X.L., and C.-Z. Chen provided theoretical support. B.Z., X.L., C.-Z. Chen, and C.-Z. Chang analyzed the data and wrote the manuscript with input from all authors.

**Competing interests:** The authors declare no competing interests.

**Data availability:** The datasets generated during and/or analyzed during this study are available from the corresponding author upon request.

**Figures and figure captions:**

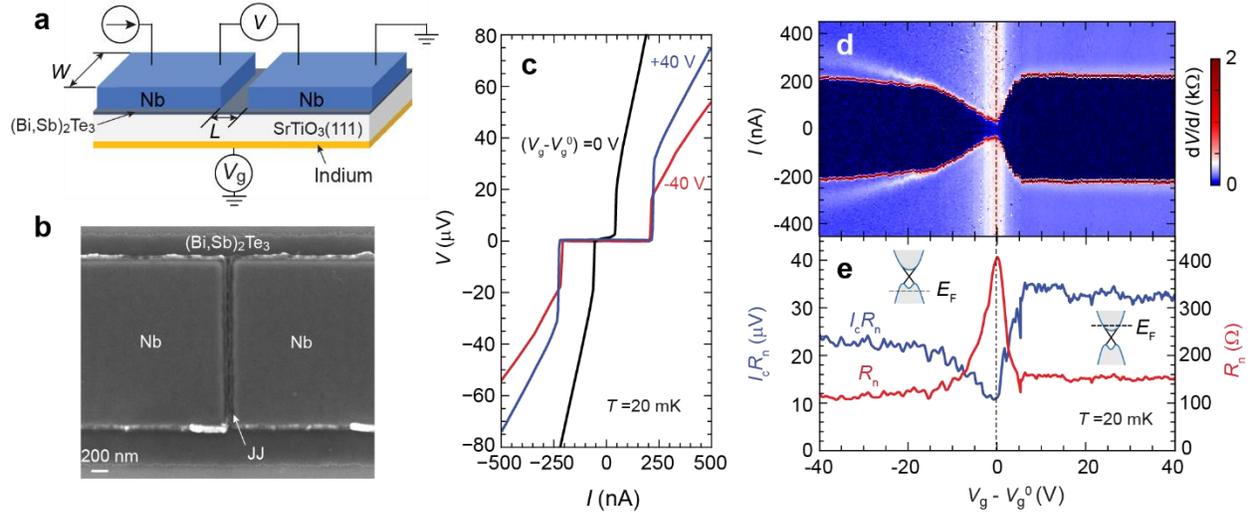

**Fig. 1| JJ-1 device on a 5 QL (Bi,Sb)$_2$Te$_3$ film. a**, Schematic of the JJ device fabricated on an MBE-grown (Bi,Sb)$_2$Te$_3$ film. **b**, The SEM image of the JJ device on a (Bi,Sb)$_2$Te$_3$ film. **c**, Three selected *V-I* curves of the JJ-1 device measured at different ($V_g$ - $V_g^0$). **d**, Color plot of differential resistance d$V$/d$I$ as a function of current bias *I* and gate bias ($V_g$ - $V_g^0$) at $B_z$ = 0 T. **e**, ($V_g$ - $V_g^0$) dependence of the normal state resistance $R_n$ (red) and the product of the critical current and the normal state resistance $I_cR_n$ (blue).

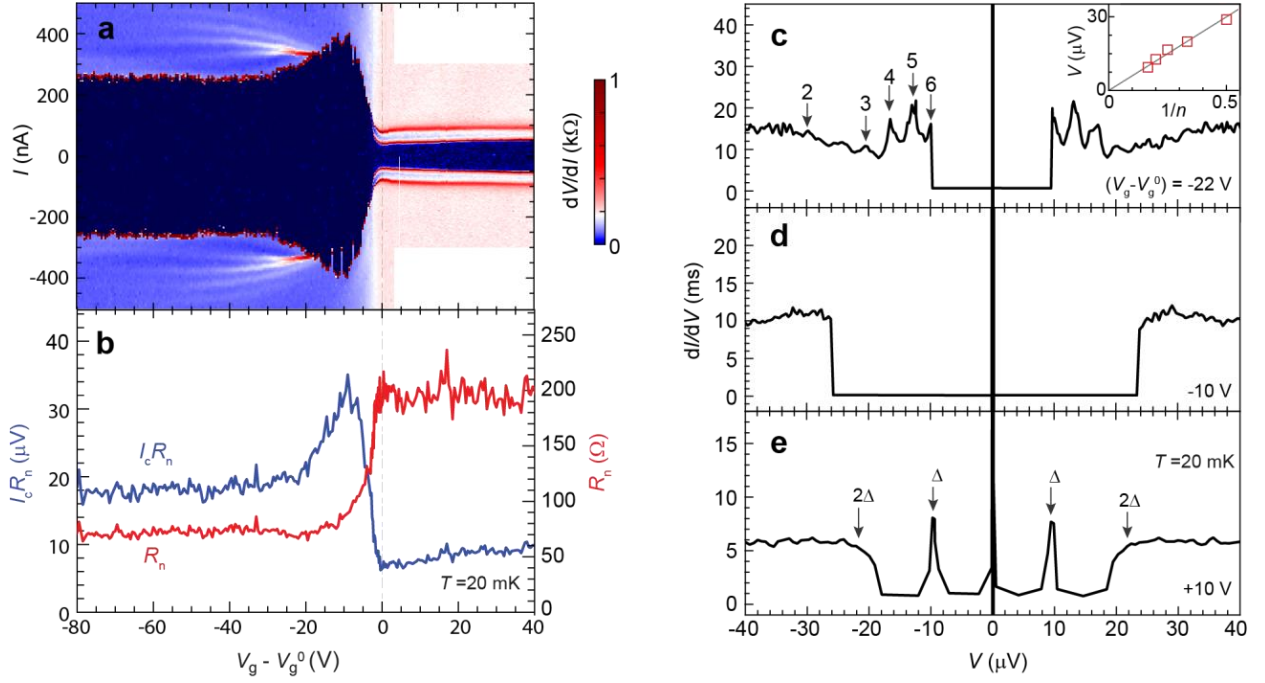

**Fig. 2| JJ-2 device on a 15 QL (Bi,Sb)$_2$Te$_3$ film. a**, Color plot of d$V$/d$I$ as a function of $I$ and ($V_g$ - $V_g^0$) at $B_z$ = 0 T. **b**, ($V_g$ - $V_g^0$) dependence of $R_n$ (red) and $I_cR_n$ (blue). **c-e**, The voltage bias $V$-dependent differential conductance d$I$/d$V$ measured at ($V_g$ - $V_g^0$) = -22 V (**c**), -10 V (**d**), and +10 V (**e**) under $B_z$ = 0 T. The index of MAR subharmonic peaks is labeled in (**c**). Inset of (**c**): MAR peak positions as a function of the peak index 1/$n$. An induced gap $\Delta$ = 30 μeV is estimated through a linear fit.

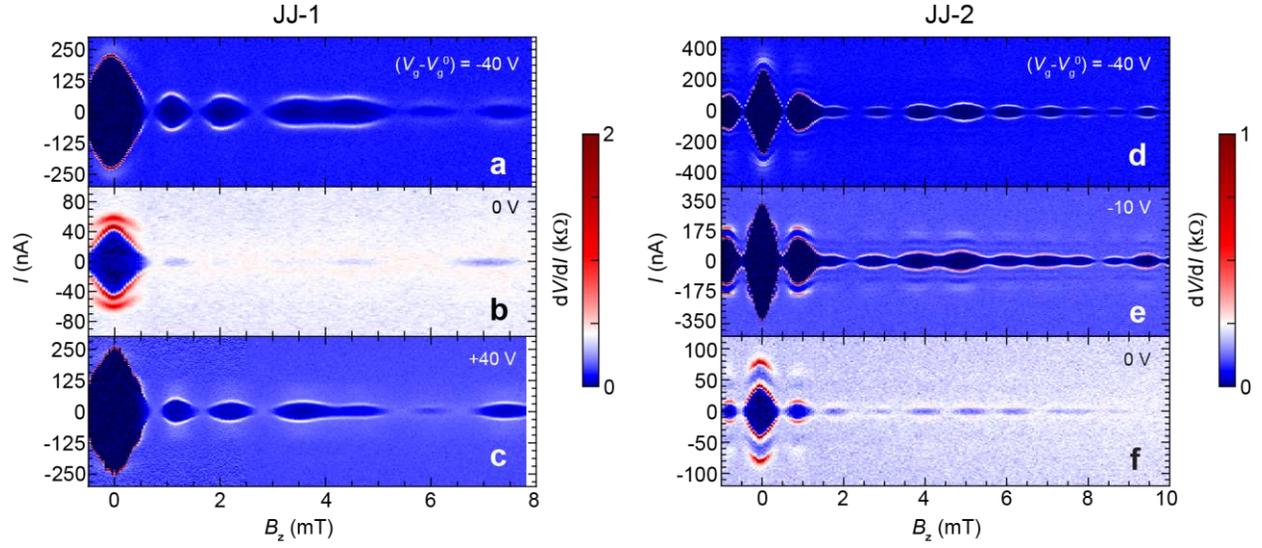

**Fig. 3| Fraunhofer patterns of the JJ-1 and JJ-2 devices. a-c,** Color plot of d$V$/d$I$ as a function of $I$ and $B_z$ at ($V_g$ - $V_g^0$) = -40 V (**a**), 0 V (**b**), and +40 V (**c**) measured on the JJ-1 device. **d-f,** Color plot of d$V$/d$I$ as a function of $I$ and $B_z$ at ($V_g$ - $V_g^0$) = -40 V (**d**), -10 V (**e**), and 0 V (**f**) measured on the JJ-2 device.

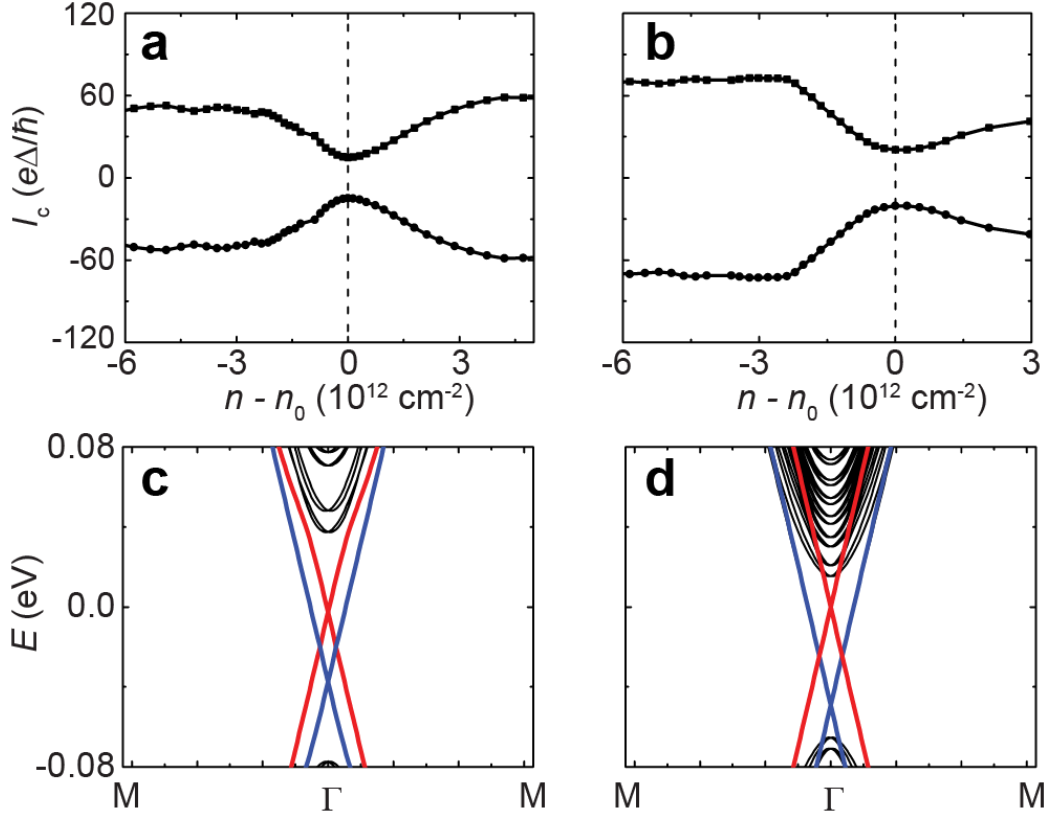

**Fig. 4| Calculated Josephson critical current in TI-based JJ devices with different *d*. a, b,** Josephson critical current $I_c$ as a function of carrier density ($n-n_0$) for the JJ devices with $d = 5$ (**a**) and $d = 15$ (**b**). The junction area is 20 nm × 2 μm. $n_0$ is the carrier density when the chemical potential lies at the top surface Dirac point of TI. **c, d,** Calculated band structures of TI films with $d = 5$ (**c**) and $d = 15$ (**d**). The bulk bands are plotted in black, while the Dirac surface states of the top and bottom layers are shown in red and blue, respectively. The red and blue lines are split because a finite potential difference shifts their bands. The potential difference is $V_0 = 0.04$ eV for $d = 5$ and $V_0 = 0.06$ eV for $d = 15$.